\documentclass[conference]{IEEEtran}
\IEEEoverridecommandlockouts
\usepackage{cite}
\usepackage{amsmath,amssymb,amsfonts}
\usepackage{algorithmic}
\usepackage{graphicx}
\usepackage{textcomp}
\usepackage{xcolor}
\usepackage{multirow}
\usepackage{booktabs}
\usepackage{hyperref}
\usepackage[table]{xcolor}
\usepackage{colortbl}
\usepackage{svg}
\def\BibTeX{{\rm B\kern-.05em{\sc i\kern-.025em b}\kern-.08em
    T\kern-.1667em\lower.7ex\hbox{E}\kern-.125emX}}
\begin{document}

\title{Targeted Test Selection Approach in Continuous Integration}
\author{
\IEEEauthorblockN{Pavel Plyusnin}
\IEEEauthorblockA{\textit{T-Technologies}\\
Email: p.plyusnin@tbank.ru}
\and
\IEEEauthorblockN{Aleksey Antonov}
\IEEEauthorblockA{\textit{T-Technologies}\\
Email: alek.antonov@tbank.ru}
\and
\IEEEauthorblockN{Vasilii Ermakov}
\IEEEauthorblockA{\textit{T-Technologies}\\
Email: v.y.ermakov@tbank.ru}
\and
\IEEEauthorblockN{Aleksandr Khaybriev}
\IEEEauthorblockA{\textit{T-Technologies}\\
Email: a.khaybriev@tbank.ru}
\and
\IEEEauthorblockN{Margarita Kikot}
\IEEEauthorblockA{\textit{T-Technologies}\\
Email: m.kikot@tbank.ru}
\and
\IEEEauthorblockN{Ilseyar Alimova}
\IEEEauthorblockA{\textit{T-Technologies}\\
Email: i.alimova@tbank.ru}
\and
\IEEEauthorblockN{Stanislav Moiseev}
\IEEEauthorblockA{\textit{T-Technologies}\\
Email: s.moiseev@tbank.ru}
}

\maketitle

\begin{abstract}
 In modern software development change-based testing plays a crucial role. However, as codebases expand and test suites grow, efficiently managing the testing process becomes increasingly challenging, especially given the high frequency of daily code commits.

 We propose Targeted Test Selection (T-TS), a machine learning approach for industrial test selection. Our key innovation is a data representation that represent commits as Bags-of-Words of changed files, incorporates cross-file and additional predictive features, and notably avoids the use of coverage maps.

 Deployed in production, T-TS was comprehensively evaluated against industry standards and recent methods using both internal and public datasets, measuring time efficiency and fault detection. On live industrial data, T-TS selects only 15\% of tests, reduces execution time by $5.9\times$, accelerates the pipeline by $5.6\times$, and detects over 95\% of test failures. The implementation is publicly available to support further research and practical adoption.
\end{abstract}

\begin{IEEEkeywords}
Regression testing, Test case selection, Machine Learning, Continuous Integration
\end{IEEEkeywords}

\section{Introduction} \label{Intro}

Continuous integration (CI) is a common and widely used software engineering development practice. 
Each CI cycle involves software testing that aims to detect potential bugs in the changed code before deploying it to production. 
One of the key tasks performed in CI testing is regression testing, where new code changes are tested within each CI cycle. 
In order to prevent bugs occurring after critical stage in CI cycle (e.g. merge to master branch in a repository) all tests anyhow relevant to changed code must be executed. 

As long as a project is being developed and handled, the amount of test cases increases proportionally. 
This necessity leads to huge computational resources consumption and regression testing becomes time-consuming. 
Hence, feedback for authors of changes is available long after tests execution had been run and further development is thus collapsed until all executed tests outcomes are known. 

A substantial body of research focuses on the problems of test selection and minimization \cite{15yoo2012regression, 16jeffrey2007improving, 13jones2003test, 12harrold1993methodology, 14chen1998new, 11black2004bi, 8vahabzadeh2018fine, 10zhang2018hybrid, 7rothermel2002empirical}. In their seminal work, Rothermel et al. \cite{rothermel2001prioritizing, 7rothermel2002empirical} introduced several techniques to prioritize test cases for regression testing based on test execution information. Specifically, they explain strategies such as ordering test cases by their total coverage of code components, ordering based on the coverage of previously uncovered components, and ordering according to the estimated ability of tests to reveal faults.

Most of the mentioned techniques rely on code coverage as the primary criterion for selecting tests potentially affected by code changes \cite{7rothermel2002empirical, FastLane}. Code elements considered in these approaches may include individual operators, method control flows, entire classes, or files. Methods such as \cite{Ekstazi, VectorCAST, Sealights.io, legunsen2017starts} aim to select all tests that could potentially be impacted by the modified code elements, thereby ensuring that all candidate failures are captured. However, this often leads to the execution of a large number of tests, even when the actual code change is minimal.

To mitigate this inefficiency, subsequent work has proposed enhancements that involve ranking tests according to their code coverage and prioritizing execution of only the most relevant ones. For instance, the "Echelon" system \cite{tr-2002-15} uses a binary representation of changed code and the coverage map to iteratively measure the intersection between the current test set and modified code blocks, thereby prioritizing test execution for each change. Greedy algorithms described in \cite{Harman_etal:Seacrhalgs2007} similarly make use of coverage tables to rank test cases within the current permutation of the test suite.

While effective in principle, relying on code coverage for test selection introduces several practical challenges in large, rapidly evolving software systems. This approach significantly reduces the number of tests that need to be run, but with large code changes affecting multiple files, there is still a significant number of tests to run. The machine learning (ML) based approach takes into account additional features and heuristics for selecting tests that cannot be considered using a coverage map. 

For example, Busjaeger and Xie~\cite{busjaeger2016learning} proposed using a machine learning model (Support Vector Machine) trained on features such as Java code coverage, textual path similarity, textual content similarity, failure history, and test age to predict the probability of test failure. According to their results, to achieve the same recall, this method required running less than 3\% of the tests—over fourteen times fewer than approaches relying solely on traditional features such as code coverage and failure history, as used in earlier non-ML-based methods.

However, code coverage maps still require a substantial amount of memory for storage; any project update necessitates rebuilding coverage information, which is both time-consuming and resource-intensive; and coverage-based approaches typically fail to account for changes in non-source code files such as configuration or resource files (e.g., \texttt{.yaml}, \texttt{.json}, \texttt{.xml})~\cite{wang2024efficient, weiss2024achieving}.

In contrast, our goal is to design approach leverages lightweight, machine learning–based methods that utilize historical test execution data for both test selection and prioritization. A key motivation is to develop less costly ML model that \textit{does not rely} on code coverage information, instead extracting informative features directly from commits. This enables the model to classify which tests are most likely to fail as a result of specific changes. However, learning optimal test case prioritization policies solely based on past test execution data can be challenging, especially for complex software systems. 

In contrast, our goal is to design approach leverages lightweight, machine learning–based methods that \textit{does not rely} on code coverage information, instead extracting informative features directly from commits for both test selection and prioritization. However, learning optimal test case prioritization policies solely based on past test execution data can be challenging, especially for complex software systems. 

\subsection{Overview of Our Development Processes}\label{overview_dev_process}

Our company manages over 40 000 repositories spanning hundreds of distinct projects. We use a Git-based version control system to manage these repositories. Repository sizes vary considerably, but for the largest repositories, there can be up to 100 commits per day. For such large and frequently updated projects, it becomes impractical to execute the full test suite after every commit. As a result, tests are accumulated throughout the day and executed in batch the so-called nightly runs. Among all types of tests, UI tests are particularly time-consuming; hence, accelerating their execution is a priority. Unless explicitly stated otherwise, we refer primarily to UI tests as "tests" in what follows.

Over time, for certain projects, test engineers and developers have built a solution\cite{shchepalin2022testimpact} (hereafter called TIA) utilizing code coverage maps and a set of rules derived from their expertise regarding which code files may impact a given test. This solution is used to determine the specific subset of UI tests to execute for each commit.

However, such coverage maps based systems are difficult to scale across the organization. For each project a custom configuration is typically required, and these systems depend on third-party code coverage tools, which are not always available for every programming language or project in the company. In addition, they require constant coverage maps that are up-to-date. Most importantly, these mechanisms often continue to select and execute a large number of tests (on one of our key projects, the TIA solution typically selects approximately 40–50\% of the total available test suite), resulting in significant test execution time and resource consumption. The unpredictable number of selected tests can also generate substantial peaks in system load.

This rapid development pace and large-scale testing introduce delays in deploying commits. Test resources, i.e., machines on which we run tests, are limited. When the number of tests to execute for a single commit is large, it can take a significant amount of time to complete all the necessary testing steps, thereby increasing feedback loop latency for developers. This motivates our problem and approach.

\subsection{Problem Statement}
Focusing on the practical value of our test selection solution, we address real-world constraints typical for large-scale industrial codebases like ours. Our repository maintains detailed metadata for every commit and records results from nightly UI test runs spanning the past 12 months. Tests are not executed per commit, but rather on aggregated daily changes, posing an additional challenge: during inference, our model makes predictions for individual changes, while training is performed on aggregated (daily) data.

Test case selection can address a wide variety of objectives. In our work, we focus on several core objectives:
\begin{itemize}
  \item \textit{Test case classification:} Predicting whether a given test case will fail for each new commit.
  \item \textit{Test case prioritization:} Ranking tests for each change by estimated failure probability.
  \item \textit{Inference time constraint:} Ensuring test selection for a change completes in under one minute (excluding model training and data preparation).
  \item \textit{Robustness to domain shift:} Maintaining accuracy despite training on aggregated (nightly) changes and inferring on individual commits.
\end{itemize}

Our problem formulation aligns more closely with Test Case Prioritization: although we solve the selection problem at the test level, we ultimately produce a ranked test list based on predicted relevance.
\subsection{Our proposal}
Assuming a complex relationship between test outcomes and source file changes, we propose a generalizable approach to test selection and prioritization. Our method analyzes each modified file and its historical association with test failures, estimating test-file relevance via hierarchical directory distance rather than coverage maps.

The contribution of this paper is as follows:
\begin{itemize}
    \item We introduce Targeted Test Selection (T-TS), a machine learning framework for industrial test selection. The core novelty is in data representation: we model commits as Bags-of-Words~\cite{Harris01081954}, incorporate cross-file features, and design additional predictive features (see Section~\ref{Features}). 
    \item We further propose an optional extension (T-TS + Code Analysis) that leverages information from code changes and test code directly.

    \item We benchmark T-TS against prior methods on open-source datasets, achieving state-of-the-art results in both speed and fault detection (see Tables~\ref{tab:exp_industrial},\ref{tab:time_tts_deeporder},~\ref{tab:exp_general_metrics}).

    \item We describe production deployment, reporting challenges and solutions. Evaluation on live data shows T-TS selects only 15\% of tests, reduces execution time by $5.9\times$, accelerates the pipeline by $5.6\times$, and detects over 95\% of test failures (see Table~\ref{prod_results}).

    \item Finally, we publicly release the implementation of T-TS.
\end{itemize}

\section{Related works}
Machalica et al.~\cite{machalica2019predictive} introduced a predictive test selection technique based on machine learning. Their model leverages test outcome history, file-level statistics (e.g., number of distinct authors, frequency of changes), and the minimal distance between changed files and tests in the dependency graph. 

We also utilize some of these features in our approach; however, several of them performed poorly in our experiments and were thus excluded from our final method. The most notable of these are discussed in Section~\ref{similar_ablation}. Furthermore, while Machalica et al. compute minimal distance using a dependency graph, our approach does not rely on such a graph. Instead, we estimate file-test proximity based on the project's directory structure (see Section~\ref{cross_files}). Additionally, whereas their method uses the raw distance value as a feature, we incorporate features derived from the closest file identified by this distance metric.

Also, a notable limitation of \cite{machalica2019predictive} approach is its reliance on the structure of an internal corporate monorepository, making it less applicable to multi-repository or open-source projects. Furthermore, both \cite{machalica2019predictive} and several other studies \cite{EALRTS, EFSM} make extensive use of code coverage data, which can be prohibitively slow to obtain for large-scale projects.

The Difference Engine \cite{ekelund2015efficient} suggested to parse and analyze results from previous test runs. It correlates code and test cases at package level and recommends test cases that are strongly correlated to recently changed packages. However, in contrast to our method, this approach relies solely on the correlation between a test and a code package. For every change, only a limited set of relevant code packages and their corresponding historical data are selected for analysis.

The FastLane approach~\cite{FastLane} focuses on a different problem: using an ML model to determine whether tests need to be run at all, i.e., whether a commit is safe. FastLane also clusters tests that often fail together, running only a representative from each cluster to minimize executions. When tests must be run, they are selected sequentially and independently of the current commit, which contrasts with our method's commit-aware selection.

In papers \cite{rl_tcp_tcs, rl_tsp, spieker2017reinforcement} reinforcement learning is used for solving test prioritization task. Although the approaches has promising results and don't use traceability links between code and test cases, in our experiments we used  it turned out to take long time for prediction, which is why we had to leave it. 

Another method, introduced in~\cite{lima2020multi}, proposes COLEMAN—an approach grounded in the Multi-Armed Bandit (MAB) framework. As MAB methods are closely related to reinforcement learning, COLEMAN inherits similar challenges, including potentially high computational complexity.

There is also the DeepOrder~\cite{sharif2021deeporder} method, which, however, employs a different set of features. Specifically, it primarily analyzes test statuses and their changes, rather than the relationship between tests and modified files, as in our approach. Furthermore, DeepOrder utilizes a neural network as its predictive model and, importantly, relies on synthetically augmented data during training—both of which distinguish it from our method.

The use of code changes and test information to predict failures appears promising, as it enables modeling of non-trivial dependencies between tests and files. For instance, \cite{jabbar2022test2vec} proposes constructing embeddings for call chains, which requires detailed trace generation. More relevant approaches analyze the code directly: \cite{meding2020early} applies Bag-of-Words (BoW) and word embeddings to code, while \cite{lousada2020neural, zhang2022comparing} learns custom embeddings for files and tests.

Our goal is to process source code directly in a way that is language-agnostic, treating code-based features as a supplement to improve test selection—unlike prior work, which often relies on language or dataset-specific techniques.

A notable distinction is that \cite{meding2020early, gao2024obtaining} applies BoW to source code itself. In contrast, our method treats each commit as a "document" and the modified files within it as "words," using BoW over sets of changed files. In this context, the unordered nature of BoW is not only natural but may better capture the semantics of code changes.



\section{Method}
\label{Method}
Our T-TS method is based on the machine learning model and a rich set of features. The primary objective is to establish a direct relationship between tests and files exclusively using machine learning techniques. We further assume that the project follows a logical directory structure organized into modules. Consequently, cross-file features are selected based on the hierarchical distance between tests and files.

In this section, we detail the data processing pipeline, the feature set, the T-TS methodology itself, and the approach used to filter out unstable tests.

\subsection{Features}
\label{Features}
Our repository structure, the initial testing environment and the accumulated logs ultimately determine the available data and its format. It is important to note that our repository organization is fairly typical, and the features we log and analyze are available on the vast majority of modern enterprise platforms~\cite{shakikhanli2025repository}.
 
The features we use can be categorized into three distinct groups: file features, test features, and cross files features.

\textbf{File features} include all files along with their historical information, 
collected as features, including:
\begin{itemize}
    \item \textit{Change flag} indicates whether the file has been modified.
    \item \textit{Number of distinct authors} represents the frequency with 
		which the file has been altered by different unique authors.
    \item \textit{Lines added/deleted} denote the total number of lines added to 
		or removed from the file.
    \item \textit{Change type} classifies the type of change, in order to the 
		descriptions provided by Git.
    \item \textit{Number of changes} reflects the frequency of modifications 
		made to the file over the past 3, 14, and 56 days, irrespective of 
		the type or size of the change. 
\end{itemize}

\textbf{Test features} provide historical information regarding:
\begin{itemize}
    \item \textit{Failure rate} is the percentage of test failures over periods of 7, 14, and 28 days.
\end{itemize}

We selected the specified time intervals for the number of changes and failure rate features because our company operates on 7-day sprint cycles. Additionally, similar intervals were used in Machalica M. et al.~\cite{machalica2019predictive}.

\label{cross_files}To assess the relevance of a candidate test to a commit, we introduce the concept of \textbf{Cross Files} - the changed files closest to the test file, where proximity is defined by directory structure distance. For each test, we consider the three nearest changed files, incorporating all their features and file extensions as described in our file data representation. To our knowledge, this feature has not been previously proposed in the literature.

Since a commit may modify multiple files, we filter out files that are rarely or frequently changed, aggregating their information into an \textbf{unknown files data} feature.

For each commit, file information is duplicated across all associated tests, while cross file and test features are computed separately for each test.

\subsection{Data Representaion} \label{data_represent}
After collecting historical data on files and tests, we encountered challenges in effectively combining this information while preserving the ability to identify the specific file responsible for a test failure. Prior work, such as that of Lousada et al.~\cite{lousada2020neural}, has employed a cross-merge approach that generates all possible test-file pairs. Although this method combines file and test statistics, it does not explicitly capture the correlations between particular tests and the files they directly interact with. Another challenge is aggregating predictions across all modified files to decide whether a test should be run for a given commit. We experimented with this representation, but the resulting predictive performance was unsatisfactory.

Consequently, in our work we adopted the $(\text{commit}, \text{test})$ representation, in which commit features also encode properties of each modified file.
\subsection{Commit as Bag-of-Words representation} \label{BoW}
A remaining challenge is the variable number of files changed in each commit, as most ML tasks assume a fixed-length feature vector. To address this, we proposed to adapt the Bag-of-Words approach \cite{Harris01081954} - popular text representation technique in natural language processing (NLP) that treats a document as a collection of words, disregarding their order and focusing on their frequency. In our setting, each commit is analogous to a document, and each modified file corresponds to a word. Thus, we represent each commit as a vector of length equal to the number of files in the repository multiplied by the number of features per file, where only the entries corresponding to the modified files in the commit have non-zero values.

Formally, let $\mathcal{F} = \{f_1, f_2, \ldots, f_N\}$ denote the set of all $N$ files in the repository, $\mathcal{F}_C \subseteq \mathcal{F}$ the subset of files modified in commit $C$, and $\mathbf{v}_{f_i} \in \mathbb{R}^d$ the feature vector for file $f_i$. The commit representation $\mathbf{x}_C \in \mathbb{R}^{N \times d}$ is constructed as:
\[
\mathbf{x}_C = \left(\mathbf{x}_{C,1}, \mathbf{x}_{C,2}, \ldots, \mathbf{x}_{C,N}\right),
\]
where
\[
\mathbf{x}_{C,i} = 
\begin{cases}
\mathbf{v}_{f_i}, & \text{if } f_i \in \mathcal{F}_C, \\
\mathbf{0}, & \text{otherwise}.
\end{cases}
\]

To the best of our knowledge, this work is the first to propose this structured sparse representation for commits in the context of CI/CD pipeline optimization. This approach explicitly models file modification patterns while maintaining compatibility with standard ML algorithms (see Figure~\ref{features_repr})

\subsection{Datasets}
We conducted experiments on four datasets: a proprietary Corporate Mobile Application (CMA) dataset collected from our internal projects and three publicly available benchmarks derived from real-world continuous integration CI environments.

The CMA dataset consists of a test suite targeting user interface functionalities within a mobile application. Unfortunately, we were unable to identify open-source datasets that include both detailed repository file information and test execution results in a format compatible with all of the feature sets used in our study. While many prior works on test selection do not provide direct comparisons on public datasets\cite{EALRTS, machalica2019predictive, busjaeger2016learning}, our goal is to deliver a comprehensive, reproducible, and large-scale evaluation against the current state-of-the-art (SOTA) methods (see Table 4 in \cite{arani2024systematic}), specifically those most closely aligned with our approach: the RETECS~\cite{spieker2017reinforcement}, COLEMAN~\cite{lima2020multi}, and DeepOrder~\cite{sharif2021deeporder} algorithms. These methods have reported results on several open industrial datasets, including Input/Output Control (IOF/ROL)~\cite{DVN/GIJ5DE_2020}, GSDTSR~\cite{google_dataset}, and OpenShift~\cite{RedHat}. Therefore, we also include these datasets in our comparison, despite the incomplete feature coverage for our method.

Table~\ref{tab:industrial_datasets} provides detailed statistics for all four datasets. It is important to note that the IOF/ROL~\cite{DVN/GIJ5DE_2020} and GSDTSR~\cite{google_dataset} datasets lack information about modified files and tests, including the number of changes and their associated paths. As a result, we were unable to construct file-level and cross features for these datasets; however, we computed test-level features, which were used for model training. The Red Hat~\cite{RedHat} dataset, by contrast, contains many of the same features employed in our approach, but it adopts a different methodology for dataset construction, specifically using cross-merge operations to combine file and test features.

\begin{table}[h]
    \centering
    \caption{Overview of open source industrial datasets used in the training and evaluation of T-TS}
    \resizebox{\linewidth}{!}{
        \begin{tabular}{@{}lccccc@{}}
            \toprule
            Data Set        & Test Cases & CI Cycles   & Verdicts & Failed \\ \midrule
            CMA &  6 580 &  55 &    64 777 &     0.8\%   \\
            Red Hat Openshift\cite{RedHat} &  5 888 &  105 &    618 312 &     10.16\%   \\
            IOF$\backslash$ROL\cite{DVN/GIJ5DE_2020}  & 2 086        & 320   & 30 319   & 28.43\%      \\ 
            GSDTSR\cite{google_dataset}   & 5 555        & 336   & 1 260 617   & 0.25\%  &       \\ 
            \bottomrule
        \end{tabular}
    }
    \label{tab:industrial_datasets}
\end{table}

\subsection{Model training}\label{model_training}
\begin{figure}[t]
	\includegraphics[width=\linewidth]{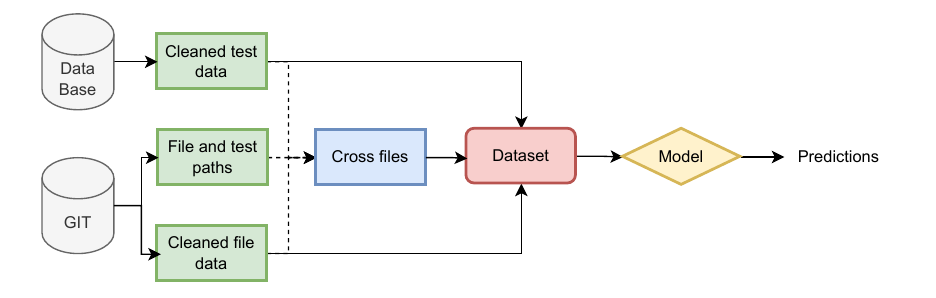}
	\caption{The training pipeline schema for the T-TS models}
	\label{fig:pipeline}
\end{figure}

Figure \ref{fig:pipeline} illustrates the complete training pipeline. 
Initially, test data was extracted from the database, including flags provided by our internal software tool based on Allure Report~\cite{Allure}. These flags identify flaky and broken tests, enabling us to filter out unstable test cases. This step is essential, as some test failures may result from causes unrelated to code changes.

Subsequently, we retrieved historical file data, as well as file and test paths, from the Git repository. This historical information allowed us to characterize both rarely and frequently modified files, while the path data was used to generate cross files features. Finally, test, file, and cross files features were merged using the commit hash as a unique key.

The resulting feature vectors were then provided to the model, which assigned each test a score between 0 and 1, representing the estimated likelihood of test failure. After calculating scores for all tests, they were ranked in descending order, and the top $N$ tests were selected for execution, where $N$ is a user-defined threshold.

We evaluated various models across different time intervals, but the best performance was achieved using a gradient-boosted decision tree classifier~\cite{chen2016xgboost} trained on data accumulated over a two-month period. Validation was conducted on data from the subsequent two weeks. Maintaining strict chronological order during data splitting was crucial to avoid data leakage.

During training, we employed binary cross-entropy as the loss function. Although the objective was to maximize the detection of failures, it was important to avoid incentivizing the model to simply predict an excessive number of failing tests. Therefore, hyperparameters were optimized by maximizing the F1 score.

\subsection{T-TS model with Code Analysis}\label{code_feat}
Our historical dataset (described in Section \ref{Features}), from the original T-TS model, relies on file metadata without leveraging code content. While this approach suffices for scenarios where code accessibility is restricted, it imposes significant limitations in most practical settings. Incorporating code semantics from test files and modified source files enables modeling more complicated dependencies. We emphasize that the proposed modification is optional—it substantially boosts predictive performance but incurs higher computational costs.

\begin{figure*}[h]
        \centering
	\includegraphics[width=0.8\linewidth]{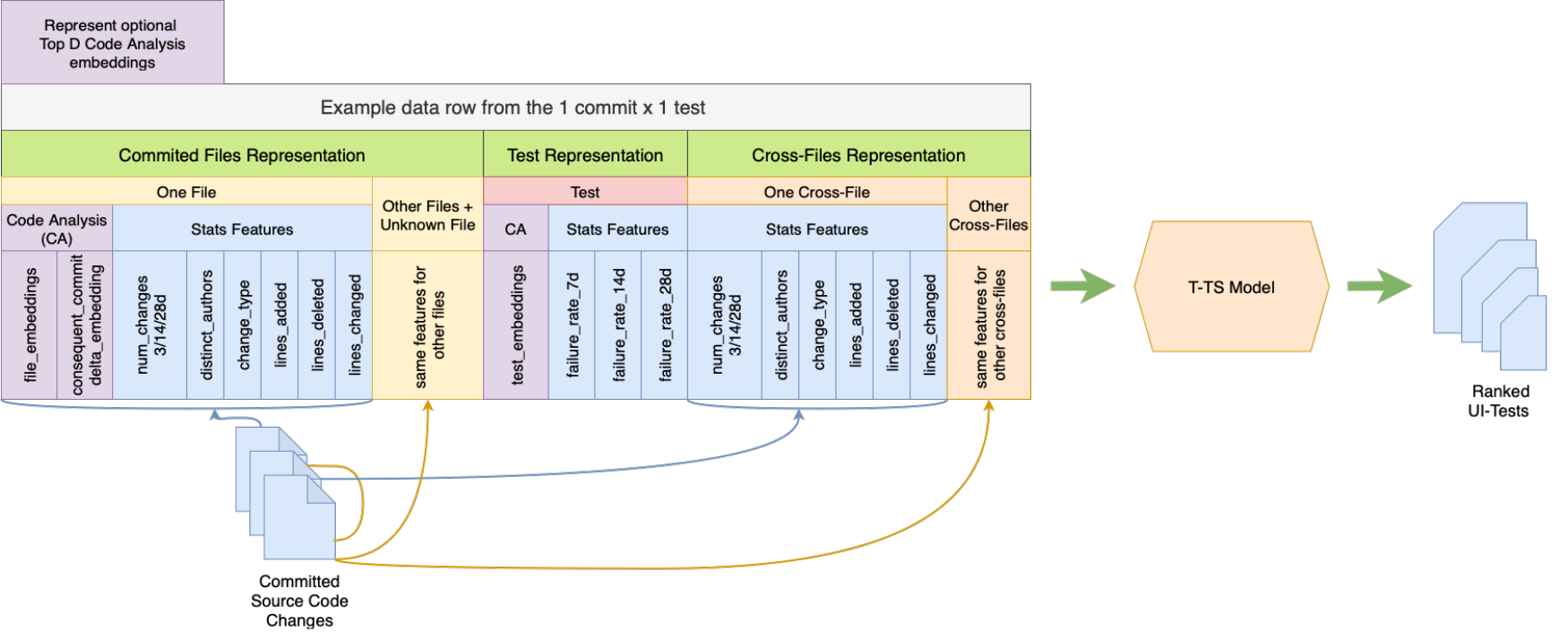}
	\caption{The data representations for T-TS and T-TS with Code Analysis}
	\label{features_repr}
        \end{figure*}

The process of building code embeddings involves several sequential steps:
\begin{itemize}
    \item \textbf{Code Parsing and Tokenization.} To extract tokens from changed files, we first split the code into two corpora—before and after the commit. Tokens are then obtained for both versions. 

    \item \textbf{Embedding Model Use.} We employ pre-trained models to obtain code embeddings for test cases, $emb_{test_i}$ for all tests in repository :$\forall i \in 1, \ldots, K$, and for each modified file both before ($emb_{f_j}^{\text{before}}$) and after ($emb_{f_j}^{\text{after}}$) the commit, for all modified files in commit $f_j \in \mathcal{F}_C$. The semantic difference in code is then modeled as the difference between embeddings: $emb_{f_j}^{\Delta} = emb_{f_j}^{\text{after}} - emb_{f_j}^{\text{before}}$. Test embeddings can be precomputed during the training phase and subsequently cached at inference time, with new embeddings generated only for newly added or modified tests during T-TS deployment without necessitating full model retraining.

    The majority of files in our dataset are written in Kotlin, making it essential for the embedding model to effectively process Kotlin code. Consequently, we selected StarEncoder~\cite{StarEncoder} and StarCoder2~\cite{StarCoder2}, both of which have been trained on Kotlin corpora. Since StarCoder2 is computationally intensive, we also experimented with a custom TF-IDF~\cite{tfidf}-based embedding model, which offered faster embedding generation but significantly lower performance. As StarCoder2 demonstrated superior performance in our experiments, we used it as the primary embedding model. 

    \item \textbf{Dataset Expansion}
    The resulting embeddings for tests, $emb_{test_i}$, as well as for each relevant file—$emb_{f_j}^{\text{after}}$ and $emb_{f_j}^{\Delta}$—are incorporated into the historical dataset (as described in Sections~\ref{Features} and~\ref{data_represent}) as additional features for tests and files, respectively.

    However, concatenating all dimensions of the test and file embeddings would be inefficient, as many coordinates may contain irrelevant information, leading to excessive and unnecessary memory usage proportional to the embedding size. To address this, we select the $D$ most important dimensions from the embedding features, where importance is assessed based on feature gain values. Through extensive experimentation, we determined that setting $D$ to the $0.1$-quantile (i.e., the top $10\%$) of features ranked by importance provides an optimal balance for enhancing the statistical dataset.

    The data representations for T-TS and T-TS with Code Analysis are illustrated in Figure~\ref{features_repr}. The subsequent use of these features and the upper-level model training process remain unchanged, as previously described in Section~\ref{model_training}.
\end{itemize}

\section{Experiments}

In this section, we describe the details of our experiments. We compare the results of our T-TS approach with those of previously proposed methods using industrial datasets from research papers. We also compare our approach to a previous one TIA that was implemented in our system on the whole test running pipeline.

\subsection{Evaluation Metrics}

As the T-TS model is a classification model, we report classification metrics, including accuracy, precision, recall, F1-score and Matthews Correlation Coefficient (MCC). Next, in order to evaluate the effectiveness of test fault detection, we present Average Percentage of Faults Detected (APFD) \cite{elbaum2002test} and Normalized Average Percentage of Faults Detected (NAPFD) \cite{qu2007combinatorial} metrics results.

To calculate the NAPFD the following formula is given:
        \[
            \text{NAPFD} = p - \frac{TF_1 + TF_2 + ... + TF_m}{m \cdot k} + \frac{p}{2k}
        \]
        
In this formula, $k$ represents the total number of tests, while $m$ denotes the number of failed tests. The variable $p$ is defined as the ratio of detected failed tests to the total number of failed tests, $m$. When all tests are selected, $p$ equals 1, making NAPFD equivalent to APFD. Lastly, $TF_i$ refers to the count of failed tests when the tests are sorted in descending order based on the predicted score; if $TF_i$ is not selected, its value is 0.



While the primary strategy for selecting tests to run involves choosing the top $k$ tests with the highest predicted scores, we also include $k$-metrics in the final report. The $k$ value, which influences both NAPFD and $k$-metrics, is based on user preferences. In our project, we selected the average number of failed tests over the past two months, which was approximately 0.8\% — equivalent to 50 failed tests for a single commit. After computing the $k$-metrics with $k$ set to 50 for each commit in the test set, we averaged the results across all commits.

\textbf{Confidence Curve:} This curve shows the cumulative proportion of failing tests detected ($y$-axis) versus the proportion of selected tests ($x$-axis) when tests are ranked by predicted failure probability. A steeper and higher curve indicates a more effective model, as more failures are detected with fewer executed tests. Comparative assessment is performed by visually and quantitatively analyzing which method’s confidence curve most quickly approaches the top left corner of the plot. 
\subsection{Results on Open-Source Industrial datasets}
        Table \ref{tab:exp_industrial} presents the results for the IOF$\backslash$ROL\cite{DVN/GIJ5DE_2020} and GSDTSR\cite{google_dataset} datasets, comparing the performance of the T-TS model against RETECS \cite{spieker2017reinforcement}, COLEMAN\cite{lima2020multi}, and DeepOrder\cite{sharif2021deeporder} algorithms. To provide a brief overview of these approaches: RETECS employs a reinforcement learning approach, COLEMAN is based on Multi-Arm Bandit, and DeepOrder uses a deep learning-based approach. The metrics used for the comparison include APFD and NAPFD, with a 50\% limit on the number of selected tests.

        Table \ref{tab:exp_industrial} displays the results for the Red Hat OpenShift\cite{RedHat} datasets, comparing them with the results obtained from the Alon Mannor's work \cite{RedHat}. In this experiment, classification metrics were utilized for evaluation.
        
        Based on the analysis of the results presented in the tables, we can draw the following conclusions:
        \begin{itemize}
            \item T-TS model demonstrates improvements in accuracy, recall, and f1-score on the Red Hat OpenShift\cite{RedHat} dataset compared with the original solution.
            \item T-TS model outperforms RETECS\cite{spieker2017reinforcement} and COLEMAN\cite{lima2020multi} algorithms on the IOF$\backslash$ROL\cite{DVN/GIJ5DE_2020} dataset, while its performance is comparable on the GSDTSR\cite{google_dataset} dataset. Additionally, since these approaches are based on reinforcement learning or are closely related to it, they exhibit higher time complexity compared to the T-TS model. 
            \item Compared to DeepOrder\cite{sharif2021deeporder}, the T-TS model shows almost identical or better results on the IOF$\backslash$ROL\cite{DVN/GIJ5DE_2020} dataset and shows significantly higher metrics on the GSDTSR\cite{google_dataset} dataset. In addition, as shown in Table \ref{tab:time_tts_deeporder}, the T-TS model performs significantly faster in all of the scenarios mentioned scenarios (see Table IV in DeepOrder\cite{sharif2021deeporder} for time metric description), achieving respectively 150X and 20X speedups in the Total algorithm running time (TT).  
        \end{itemize}

        \begin{table}[h]
            \caption{Comparison with RETECS\cite{spieker2017reinforcement}, COLEMAN\cite{lima2020multi}, and DeepOrder\cite{sharif2021deeporder} methods on the industrial open-source datasets}
            \centering
            \resizebox{\linewidth}{!}
            {%
                \begin{tabular}{l | rrrr | rr | rr }
                    \toprule
                    & \multicolumn{4}{c|}{Red Hat Open Shift\cite{RedHat}} & \multicolumn{2}{c|}{IOF$\backslash$ROL\cite{DVN/GIJ5DE_2020}} & \multicolumn{2}{c}{GSDTSR\cite{google_dataset}}\\
                    Method & Accuracy & Precision & Recall & F1-score & APFD & NAPFD & APFD & NAPFD \\ \midrule
                    RETECS\cite{spieker2017reinforcement} & - & - & - & - & 0.50 & 0.50 & \textbf{0.99} & \textbf{0.99} \\
                    COLEMAN\cite{lima2020multi} & - & - & - & - & 0.51 & 0.51 & \textbf{0.99} & \textbf{0.99} \\
                    DeepOrder\cite{sharif2021deeporder} & - & - & - & - & \textbf{0.67} & 0.56 & 0.94 & 0.79 \\
                    Red Hat\cite{RedHat} & 0.95 & \textbf{0.77} & 0.33 & 0.47 & - & - & - & - \\\midrule
                    T-TS & \textbf{0.96}& 0.63& \textbf{0.64}& \textbf{0.63}& \textbf{0.67}& \textbf{0.60}& \textbf{0.99}& \textbf{0.99}\\
                    \bottomrule
                \end{tabular}
            }    
            \label{tab:exp_industrial}
        \end{table}

\begin{table}[h]
    \centering
        \caption{Time costs for T-TS and DeepOrder (in seconds)}
        \begin{tabular}{@{}l|cc|cc@{}}
            \toprule
              & \multicolumn{2}{c|}{GSDTSR} & \multicolumn{2}{c}{IOF/ROL}\\
              Metric & T-TS & DeepOrder & T-TS & DeepOrder\\ \midrule
              FT & \textbf{0.126} & 120.000 & \textbf{0.008} & 3.021 \\ \midrule
              LT & \textbf{0.161} & 437.489 & \textbf{0.009} & 22.427 \\ \midrule
              AT & \textbf{0.143} & 58.651 & \textbf{0.008} & 1.522 \\ \midrule
              PT & \textbf{34.362} & 39.000 & \textbf{0.238} & 14.000 \\ \midrule
              RT & \textbf{0.125} & 13.000 & \textbf{0.008} & 0.263 \\ \midrule
              TT & \textbf{34.522} & 716.000 & \textbf{0.247} & 38.000 \\ \midrule
        \end{tabular}
    \label{tab:time_tts_deeporder}
\end{table}

\subsection{Results on the internal proprietary CMA dataset}
Table~\ref{tab:exp_general_metrics} summarizes the results of our experiments with the \textit{T-TS} and \textit{T-TS + Code Analysis} models and baselines, each trained on two months of historical data and evaluated over the subsequent two weeks on the CMA dataset.

\begin{figure}[t]
            \
            \includegraphics[width=\linewidth]{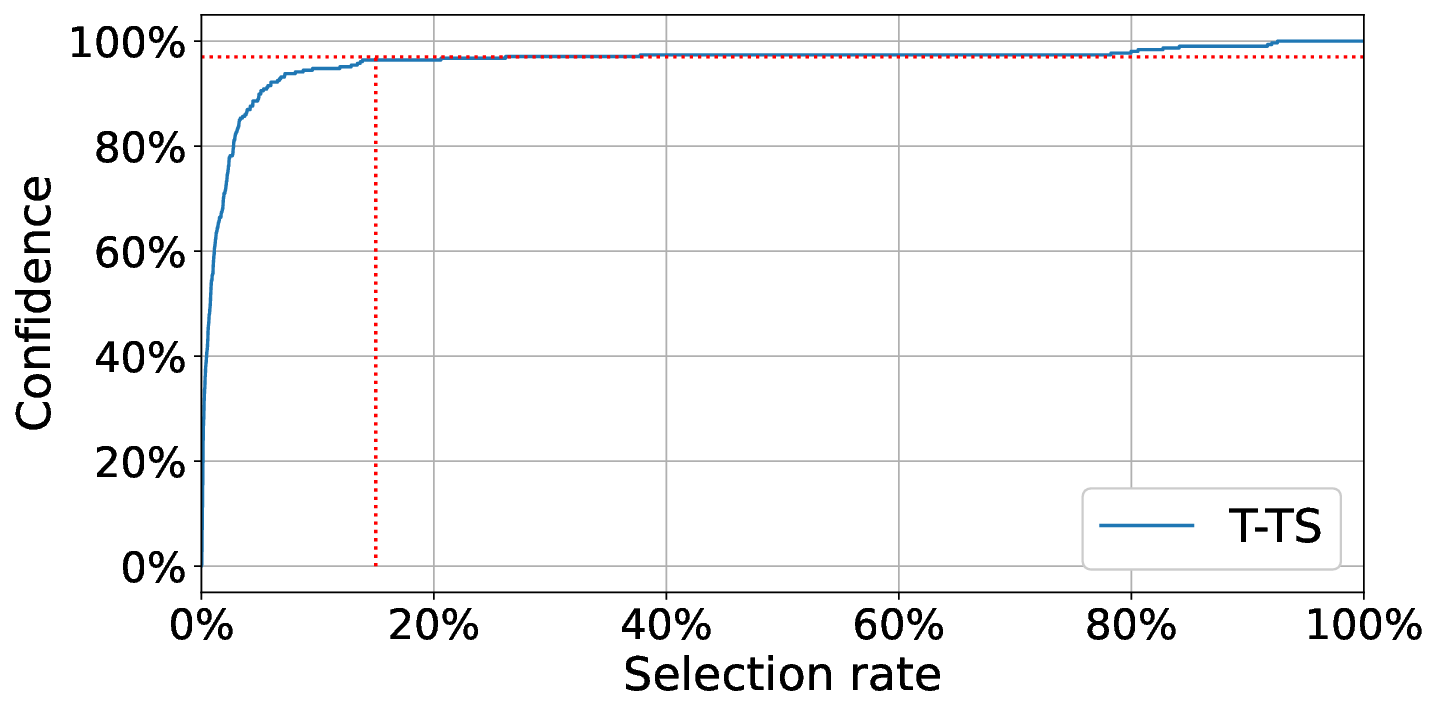}
            \caption{Confidence curve for T-TS method. Here, the X axis shows the proportion of tests run, and the Y axis shows the proportion of the found failed tests}
            \label{fig:exp_general_confidence}
\end{figure}

\begin{figure}[h]
	\includegraphics[width=\linewidth]{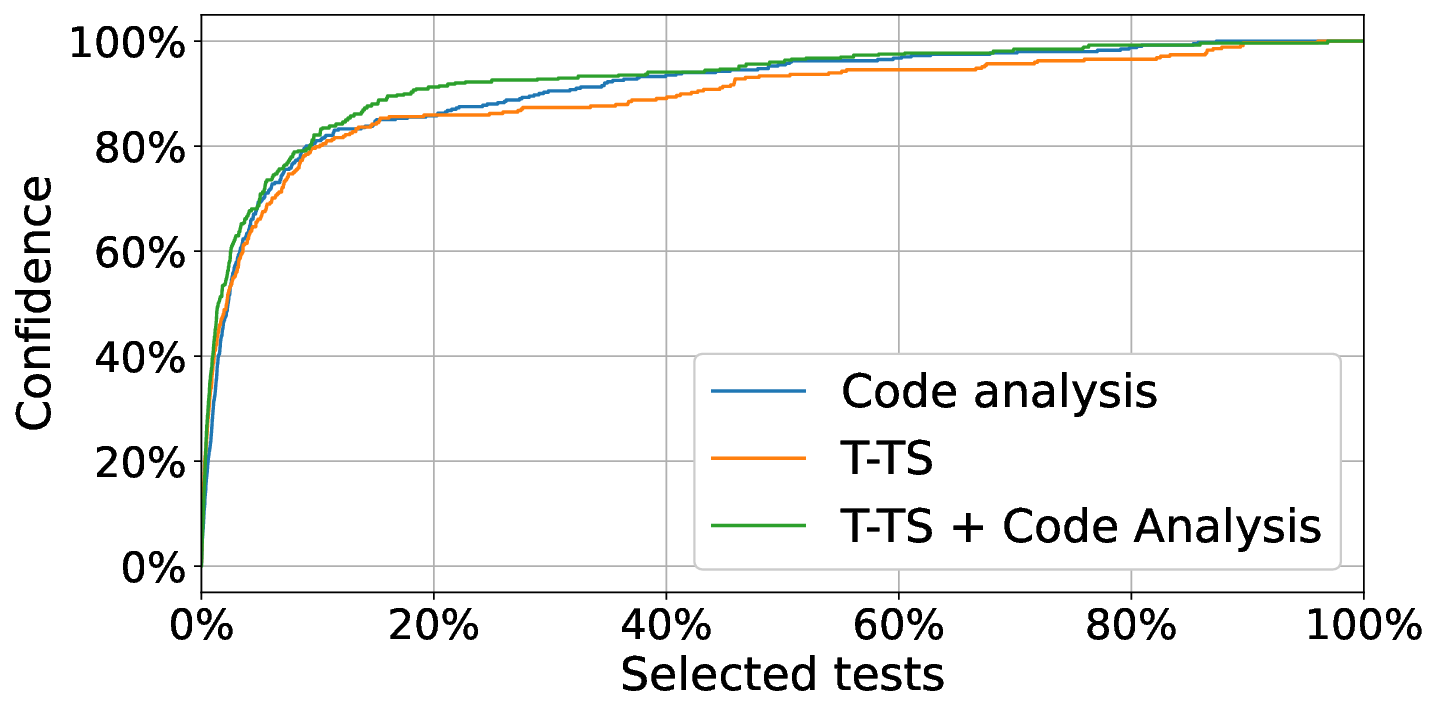}
	\caption{Confidence Curve comparison for Code Analysis approaches}
	\label{cba_combination_confidence}
        \end{figure}

Figures~\ref{fig:exp_general_confidence} and~\ref{cba_combination_confidence} present the corresponding confidence curves, illustrating the relationship between the proportion of selected tests and the proportion of detected failed tests.

To evaluate the practicality of deploying these methods in real CI cycles, all experiments were performed without GPUs on a machine with a 16-core Intel(R) Xeon(R) Gold 5320 CPU @ 2.20GHz and 320 GiB RAM.

Based on the data in Table~\ref{tab:exp_general_metrics} and the confidence curves, we highlight the following key findings:

\begin{itemize}
    \item Selecting 15\% of all tests using the T-TS model is sufficient to detect 97\% of all failed tests.
    \item Selecting only 0.8\% of all tests—which corresponds to the average proportion of failing tests—using the T-TS model enables the detection of 60\% of all failed tests, with a NAPFD of 0.47 and an F1-score@K of 0.30.

     \item The more advanced \textit{T-TS + Code Analysis} approach, which combines historical data with the top $D$ most important dimensions of test case and modified source file code embeddings, results in substantial improvements across all classification and fault detection metrics. However, the time and computational resources required for the full pipeline—including extraction of source file statistics and selection of embedding dimensions—do not currently meet our product requirements for practical industrial deployment.

\end{itemize}

\begin{table*}[h]
        \caption{Overall evaluation results of proposed T-TS methods and modifications on CMA dataset}
        \centering
        \resizebox{\textwidth}{!}{
            \begin{tabular}{l | l | ccccc | ccc | cc | c}
                \toprule
                    && \multicolumn{5}{c|}{Classification metrics} & \multicolumn{3}{c|}{Average $k$-metrics} & \multicolumn{2}{c|}{Fault detection metrics} & Inference \\
                    Experiment type & Method & Accuracy & Precision & Recall & F1-score & MCC & Precision@50 & Recall@50 & F1-score@50 & APFD & NAPFD & time, sec\\
                    \midrule
                    \multirow{2}{*}{Baseline}
                    & Random k=50 tests & \textbf{0.99} & 0.004  & 0.0008 & 0.001 & 0.001 & 0.004 & 0.0008 & 0.005 & 0.02 & 0.0003 & 0.95\\
                    & TIA\cite{shchepalin2022testimpact} & 0.96 & 0.01 & 0.28 & 0.02 & 0.03 & — & — & — & — & — & 121.43   \\
                    \midrule
                    \multirow{2}{*}{Main Methods}
                    & T-TS & \textbf{0.99} & 0.55 & 0.39 & 0.46 & 0.47 & 0.23 & 0.47 & 0.30 & 0.95 & 0.47 & 22.12 \\ 
                    & T-TS + Code Analysis & \textbf{0.99} & \textbf{0.73} & \textbf{0.72} & \textbf{0.72} & \textbf{0.72} & \textbf{0.62} & \textbf{0.82} & \textbf{0.70} & \textbf{0.98} & \textbf{0.51} & 420.21 \\
                    \midrule
                    \rowcolor{gray!20}
                    \multicolumn{13}{c}{\textbf{Ablation Study}} \\
                    \midrule
                    \multirow{3}{*}{File and test closeness}
                    & T-TS + Name Similarity & 0.98 & 0.34 & 0.40 & 0.37 & 0.37 & 0.42 & 0.30 & 0.35 & 0.90 & 0.17 & 23.43\\
                    & T-TS + Path Similarity & \textbf{0.99} & 0.46 & 0.41 & 0.44 & 0.43 & 0.22 & 0.56 & 0.29 & 0.97 & 0.42 & \textbf{21.97} \\
                    & T-TS + Attention Similarity& \textbf{0.99} & 0.52 & 0.31 & 0.38 & 0.40 & 0.28 & 0.47 & 0.34 & 0.97 & 0.47 & 26.63 \\
                    \midrule
                    \multirow{2}{*}{Code Analysis}
                    & Code Analysis & 0.98 & 0.21 & 0.33 & 0.26 & 0.26 & 0.24 & 0.33 & 0.28 & 0.91 & 0.19 & 120.32 \\
                    & T-TS + Code Similarity & \textbf{0.99} & 0.64 & 0.63 & 0.63 & 0.63 & 0.25 & 0.78 & 0.36 & 0.95 & 0.14 & 341.57 \\

                \bottomrule
            \end{tabular}
            }
        \label{tab:exp_general_metrics}
        \end{table*}

\section{Ablation Study}

In this section, we evaluate the effectiveness of the features proposed for our T-TS predictive models and compare their performance against alternatives. 

    \subsection{Effect of test, file, and cross features}
    \label{importanceAnalysis}
        To assess the impact of \textit{file}, \textit{test}, and \textit{cross-file} features discussed in Section~\ref{Method}, we employed several analytical methods.

        First, we applied the T-TS approach to the CMA dataset and used XGBoost’s~\cite{chen2016xgboost} built-in feature importance analysis with the \textit{'gain'} criterion. For clarity, we averaged the importance scores within each feature group. Results show that \textit{test features} have the strongest influence on predictions, although \textit{file} and \textit{cross-file} features also contribute meaningfully.

        Second, to further evaluate the significance of \textit{file} and \textit{cross-file} features, we compared model performance using the full feature set with versions excluding each group. We quantified the effects by tracking changes in NAPFD and K-metric values. As shown in Table~\ref{tab:napfd_k_metrics}, omitting \textit{file} and \textit{cross-file} features results in decreased performance on both metrics.
    

        \begin{table}[h]
            \centering
                \caption{Change in average K-metrics and NAPFD with the exclusion of
    file features or cross features}
            \resizebox{\linewidth}{!}{
                \begin{tabular}{@{}lccccc@{}}
                    \toprule
                    Approach  & NAPFD & Precision@50   & Recall@50 & F1-score@50 \\ \midrule
                    All features &  \textbf{0.472} &  \textbf{0.229} &    \textbf{0.473} &     \textbf{0.298}   \\
                    Excluding file features &  0.459 &  0.224 &    0.459 &     0.291   \\
                    Excluding cross features  & 0.459   & 0.224   & 0.460   & 0.290      \\ 
                    \bottomrule
                \end{tabular}
            }
            \label{tab:napfd_k_metrics}
        \end{table}

\subsection{Effect of Similarities features} \label{similar_ablation}
We hypothesized that file names similar to test names might enhance test selection beyond the features described in Section~\ref{Features}. To investigate this, we computed cosine similarity between test and file embeddings using both tokenized file names and full paths.

We evaluated multiple embedding methods—Word2Vec~\cite{w2v}, GloVe~\cite{glove}, and TF-IDF~\cite{tfidf}—with TF-IDF providing the best results for our project's unique abbreviations and naming conventions. Additionally, following~\cite{zhang2022comparing}, we processed tokenized names through a shared embedding layer, followed by a Bidirectional GRU~\cite{cho2014learning} and multi-head attention. The resulting encodings were used to compute a connection score via a two-layer fully connected network with sigmoid activation.

\begin{figure}[h]
	\includegraphics[width=\linewidth]{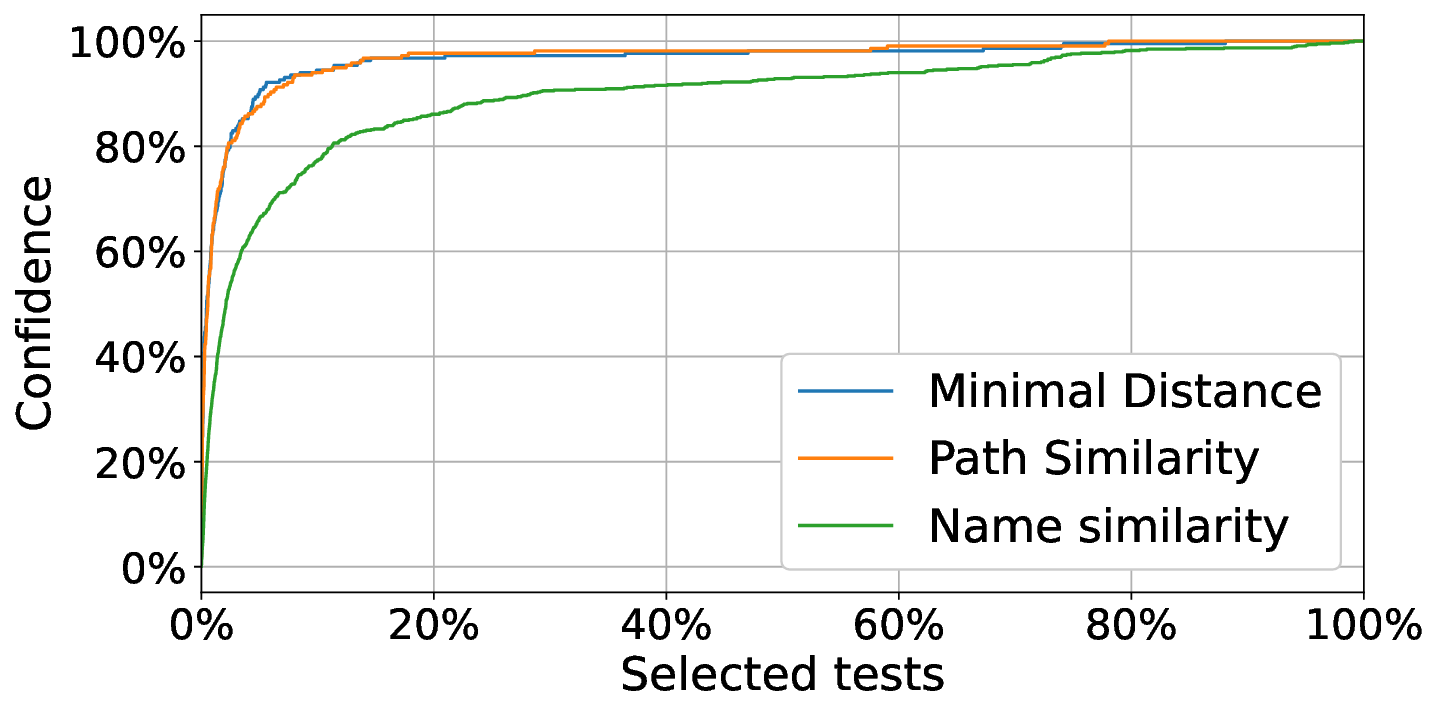}
	\caption{Confidence curve comparison of T-TS, T-TS + Name Similarity, and T-TS + Path Similarity methods.}
	\label{cross_files_by_names}
\end{figure}

Figure~\ref{cross_files_by_names} shows confidence curves for three configurations: the primary T-TS approach, and variants with Name- or Path-similarity features. Metrics are summarized in Table~\ref{tab:exp_general_metrics}.

Results indicate that including name similarity cross-features significantly degrades performance. We attribute this to:
\begin{itemize}
    \item A lack of true correspondence between similar names and actual test-file dependencies.
    \item Frequent typos, abbreviations, and inconsistencies in CI naming, which harm embedding quality.
    \item Overall, similarity features introduce noise and reduce model effectiveness.
\end{itemize}

In conclusion, hierarchical distance features in the T-TS approach encode dependencies more effectively than name-based similarity, providing a better trade-off between predictive power and evaluation speed.
    
    \subsection{Effect of Code Analysis}

    In addition to our main experiment, we evaluated whether only Code Analysis features (described in Section~\ref{code_feat}), derived solely from code analysis (without historical data), could effectively select relevant tests.

    Experiments (see the Code Analysis row in Table~\ref{tab:exp_general_metrics} and confidence curve on Figure~\ref{cba_combination_confidence}) demonstrate that this embedding-only approach performs comparably to T-TS, which relies exclusively on historical features. Notably, both datasets were constructed using the same set of commits, and XGBoost~\cite{chen2016xgboost} was used as the top-level model.

    These results show that, despite the high-level abstraction of UI test code, code-based features still provide substantial information for test selection. This finding suggests that combining historical and code-based features could further enhance the effectiveness of our T-TS approach

\section{Software Issues and Deployment Challenges}
During deployment, we encountered engineering challenges impacting both pipeline functionality and integration. Here, we outline the principal issues and mitigation strategies.

\subsection{Modular Structure and Resource Constraints}\label{modular_filtering}
In one of the target projects, the test suites are executed on a limited number of emulators. To maintain clarity and facilitate development, the team decided to split the codebase into multiple modules, each consisting of logically related components. However, the exact number of modules can vary, and it is not precisely defined at this stage. 


When a test is selected, its corresponding module must be compiled. Although unmodified modules are cached, compiling and requesting modules introduces considerable overhead, lengthening the testing pipeline. 

To address infrastructure constraints, we restrict test execution to modified modules and their closest dependencies in the hierarchy, minimizing unnecessary runs in unrelated modules. The modular subdivision is guided primarily by the presence of \texttt{gradle} files in the directories or subdirectories. These files mark functional boundaries, allowing us to define which modules need to be recompiled and tested. While this strategy may exclude some tests that could fail, On the other hand, it is critical to operate within the project's infrastructure constraints. 

\subsection{Challenges in Timing, Compilation, and Repeated Retries}

Although Targeted Test Selection (T-TS) runs fewer tests than previously adopted method based on the code coverage maps (TIA), our initial measurements revealed that T-TS pipelines still took significantly longer time to complete. Upon closer examination, we identified two main reasons for this discrepancy:

\begin{enumerate}
    \item \textbf{Repeated Retries of Failing Tests.} 
    All failing tests are typically rerun multiple times to confirm the validity of their results, aiming to eliminate sporadic or environment-specific failures. Because our predictive approach may execute a greater number of failing tests compared to previously adopted method TIA, this can lead to a substantial increase in overall test runtime. 

    \item \textbf{Compilation Overhead and Suboptimal Parallelization.} 
    Test modules do not always run in parallel due to resource constraints, and additional time is spent compiling those modules before test execution. So the running time of the whole pipelines depends not only on the number of selected tests, but also on the number of requested modules. Also, the number of available emulators is limited, so parallelism is bounded by this upper limit.
\end{enumerate}

We continually refine our time measurement methodology. Currently, we record overall test runtime in a parallel environment, excluding repeated reruns and compilation times for greater clarity on actual testing overhead. Our objective is to develop a comprehensive metric that accurately reflects the total cost of T-TS framework.

In conclusion, while T-TS reduces the number of executed tests, further efficiency gains can be achieved by more selective module filtering. Additionally, test pipeline timing should be measured based on actual test execution and infrastructure usage, rather than including developer-induced delays from manual test restarts.

\subsection{Implementation and System Design}

\begin{figure*}[t]
    \centering
	\includegraphics[width=0.8\linewidth]{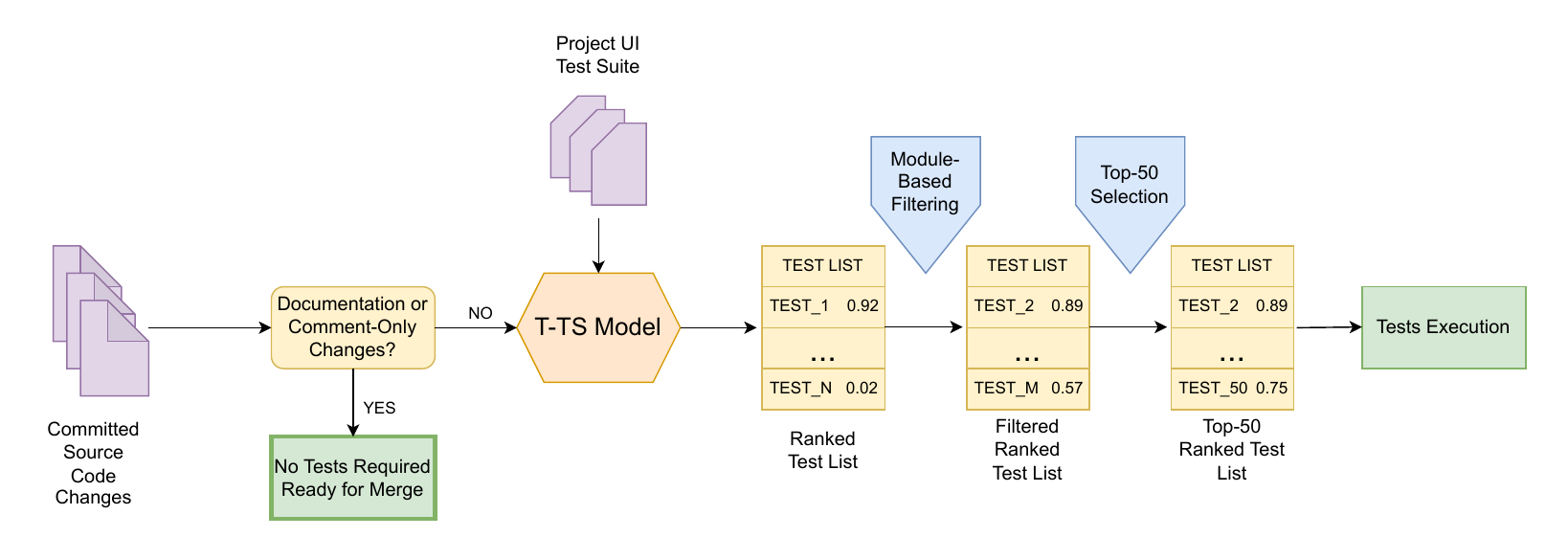}
	\caption{Integration of the T-TS framework into the CI/CD regression testing pipeline}
	\label{CI_CD_flow}
\end{figure*}

Figure~\ref{CI_CD_flow} shows a flow-graph that captures the way we
integrate our T-TS approach to the CI/CD regression testing workflow. The process can broadly be divided into several stages. 

\textbf{Data Collection and Model Retraining:} Dedicated scripts are provided to QA specialists, who periodically aggregate historical nightly runs data over a defined timeframe. This aggregated dataset is then used to train the predictive T-TS model. Although experiments show that model retraining can be deferred for up to six months without significant performance loss, in some cases—such as the active development of new modules—the lack of training data for new files can degrade performance. To ensure reliability, we retrain the model every two weeks at night, which maintains consistently high quality across projects.

\textbf{T-TS Framework Usage:}
When a developer submits a commit, T-TS checks whether any program files have been modified. If only documentation files (e.g., with the .md extension) were changed, no tests are run. While files in formats such as .txt or .yaml do not contain code, they often store configuration parameters that can impact program behavior and test outcomes. As a result, file extensions are carefully curated, and only those guaranteed not to affect functionality under any circumstances are excluded from test triggers.

Additionally, we exclude commits that alter only code comments. This check is implemented via regular expressions applied to Git diffs and is optimized for speed. The comment filter targets common patterns and is supported only for Java and Kotlin, the primary languages used in our company.

Next, we apply our T-TS model, which produces a ranked list of all project tests based on their predicted probability of failure. Given that the project encompasses numerous modules and tests, each test is associated with a specific \textit{project module}. After obtaining the ranked list of tests, a \textit{modular filtering} step, as described in Section~\ref{modular_filtering}, is performed to exclude tests from modules unrelated to the recent code change, thus reducing the scope of the test and execution time.

Each project can be configured with a specific number of tests to execute or may support sequential execution until a predefined time limit is reached. In our deployment, however, we consistently run up to 50 tests per commit (representing 0.8\% of the entire test suite). Fewer than 50 tests may be executed if, after filtering, fewer eligible tests remain.

Before tests are actually executed, they undergo an additional filtering step based on their historical stability. An internal service maintains a record of each test's historical results. Tests identified as unstable (encompassing both flaky and broken tests) are filtered out, ensuring that only stable, informative tests are executed. This final selection step further refines the test set, increasing the reliability and usefulness of the test feedback.

If no test failures are detected, the commit can be merged into the main development branch. It is important to note that nightly test runs are still performed. This practice increases reliability, enables ongoing monitoring, and helps mitigate risks associated with potential failures.

Also, T-TS system performance and metrics are continuously monitored using a dedicated Grafana~\cite{grafana} dashboard. For configurable time ranges (defaulting to the past week), the dashboard displays daily-averaged precision, recall, the proportion of failing tests among executed tests, test execution time per commit, and the number of failing tests detected in nightly runs.

\subsection{The results of industrial T-TS deployment}

We have successfully implemented the proposed T-TS solution in our company's CI/CD workflow. The Table~\ref{prod_results} shows the results of the industrial deployment on the project from which the CMA dataset was built. All values in the Table~\ref{prod_results} are averaged over a month. As you can see, the real metrics after deployment are quite correlated with the metrics obtained on the CMA dataset. By sampling only 0.8\% of the total number of tests we find more fault tests than the previously existed solution TIA\cite{shchepalin2022testimpact} based on coverage map analysis, which runs on average 32\% of all tests. So far, we have achieved almost 30-x speedup over a full pipeline run, and 2.8x speedup over the previously used solution TIA.

\begin{table}[tbh]
    \caption{The results of model deployment in corporate CI/CD workflow}
    \centering
    \resizebox{\linewidth}{!}
    {%
        \begin{tabular}{r | rrr | rr}
            \toprule
              & \multicolumn{3}{c}{Product metrics} & \multicolumn{2}{c}{Time Measurement}\\
            \cmidrule(lr){2-4}\cmidrule(lr){5-6}
            Approach & Selection Rate & Precision & Recall & Tests Run, min & Full Pipeline, min \\
            \midrule
            All Tests & 1.00 & - & - & 202.0 & 202.0 \\
            TIA  & 0.43 & 0.01 & 0.21 & 17.7 & 19.5\\
            T-TS on 0.8\% & \textbf{0.01} & \textbf{0.59} & \textbf{0.26} &  \textbf{4.7} &  \textbf{6.9}\\
            T-TS on 15\% & 0.15 & 0.02 & 0.95 &  34.0 &  36.2\\
            \bottomrule
        \end{tabular}%
    }
    \label{prod_results}
\end{table}

\section{Threads to Validity}
\subsection{Threats to External Validity}
The core principles underlying our methodology—commit risk prediction, the correlation between failing tests and code changes, and the employed feature set—are fundamental to many large-scale software development processes~\cite{shakikhanli2025repository, rahman2013and, vahabzadeh2018fine}. Moreover, our approach utilizes generic machine learning techniques and features that are not tailored to any specific properties or characteristics of our data~\cite{busjaeger2016learning, machalica2019predictive, sharif2021deeporder}. Thus, we believe our method is universally applicable to any large-scale service employing a CI/CD pipeline, provided that a sufficiently rich history of tests and commits—with relevant features—is available.

\subsection{Threats to Internal Validity}
We acknowledge an inherent trade-off between maintaining high precision and maximizing time savings. To address this, we enable QA specialists to adjust various parameters of our system, facilitating a balance between time saved and the desired level of precision. For example, as shown in Table~\ref{prod_results}, increasing the proportion of selected tests from $0.8\%$ to $15\%$ raises the detection rate of failing tests from $26\%$ to $95\%$, nevertheless with a $5.2\times$ increase in overall pipeline execution time—which remains $5.6\times$ faster than running the full test suite.

We further recognize that our pipeline is primarily intended to accelerate developer feedback and thereby speed up the development process. Any missed failing tests will still be detected during nightly test runs. Additionally, all T-TS metrics are continuously monitored, and alerting mechanisms are in place for prompt notification in case of anomalies.

\section{Conclusion}
We introduced T-TS, a novel machine learning approach for test case selection in CI/CD workflows. T-TS leverages a rich set of features, with a particular focus on modeling interactions between changed files and tests, and offers flexibility to balance between higher fault detection rate and faster execution. Production deployment accelerated our testing process by $2.8\times$. Comprehensive evaluation on both internal and public datasets demonstrated that T-TS achieves state-of-the-art results in fault detection and inference speed. The implementation is released as open source (\href{https://github.com/trndcenter/t-ts-benchmark/tree/ICSME}{https://github.com/trndcenter/t-ts-benchmark/tree/ICSME}).

\bibliographystyle{ieeetr}
\bibliography{bibliography}

\end{document}